\documentclass[cup6b]{cupbook}
\usepackage{psfig,latexsym,longtable,lscape}

\newcommand{\D}{$^\circ$}
\def\p0{\phantom{0}}
\newcommand\approxlt{\mbox{$^{<}\hspace{-0.24cm}_{\sim}$}}
\newcommand\approxgt{\mbox{$^{>}\hspace{-0.24cm}_{\sim}$}}
\def\it{\sl}

\def\arcsec{\hbox{$^{\prime\prime}$}}
\DeclareRobustCommand{\ion}[2]{%
\relax\ifmmode
\ifx\testbx\f@series
{\mathbf{#1\,\mathsc{#2}}}\else
{\mathrm{#1\,\mathsc{#2}}}\fi
\else\textup{#1\,{\mdseries\textsc{#2}}}%
\fi}

\begin{document}

\pagenumbering{roman}
\cleardoublepage

\pagenumbering{arabic}

\setcounter{chapter}{10}

\setcounter{table}{0}

\author[P. Kahabka]{P. KAHABKA\\Sternwarte, Universit\"at Bonn, 
Auf dem H\"ugel 71, D-53121 Bonn}

\chapter{Super Soft Sources}


\section{Introduction}
Super Soft Sources (SSS) are characterized by radiation with effective 
temperatures of $\sim$$10^{5}$ to $\sim$$10^{6}$~K and luminosities of 
$\sim$$10^{36}$ to $\sim$$10^{38}\ {\rm erg}\ {\rm s^{-1}}$. The first 
luminous SSS in the Large Magellanic Cloud (LMC) have been discovered around 
1980 with the {\sl Einstein} observatory (Long et al. 1981). 
Optical identifications for both CAL~83 and CAL~87 established the binary 
nature of these sources (Cowley et al. 1990; Smale et al. 1988). During  the 
{\sl ROSAT} all-sky survey in 1990--1991 in soft X-rays (cf. Tr\"umper et al. 
1991) and pointed follow-up observations many SSS have been detected.
{\sl BeppoSAX} discovered super-soft X-ray emission from recurrent and 
classical novae. Deep observations performed with {\sl Chandra} and
{\sl XMM-Newton} led to the discovery of SSS in nearby spiral and elliptical 
galaxies. A catalog of SSS is given in Greiner (2000a). Optical 
identifications exist for SSS in the Galaxy, the LMC, and the SMC
(cf. Tab.~11.1).

\begin{figure*}
  \vbox{\hskip -\leftskip
  \psfig{figure=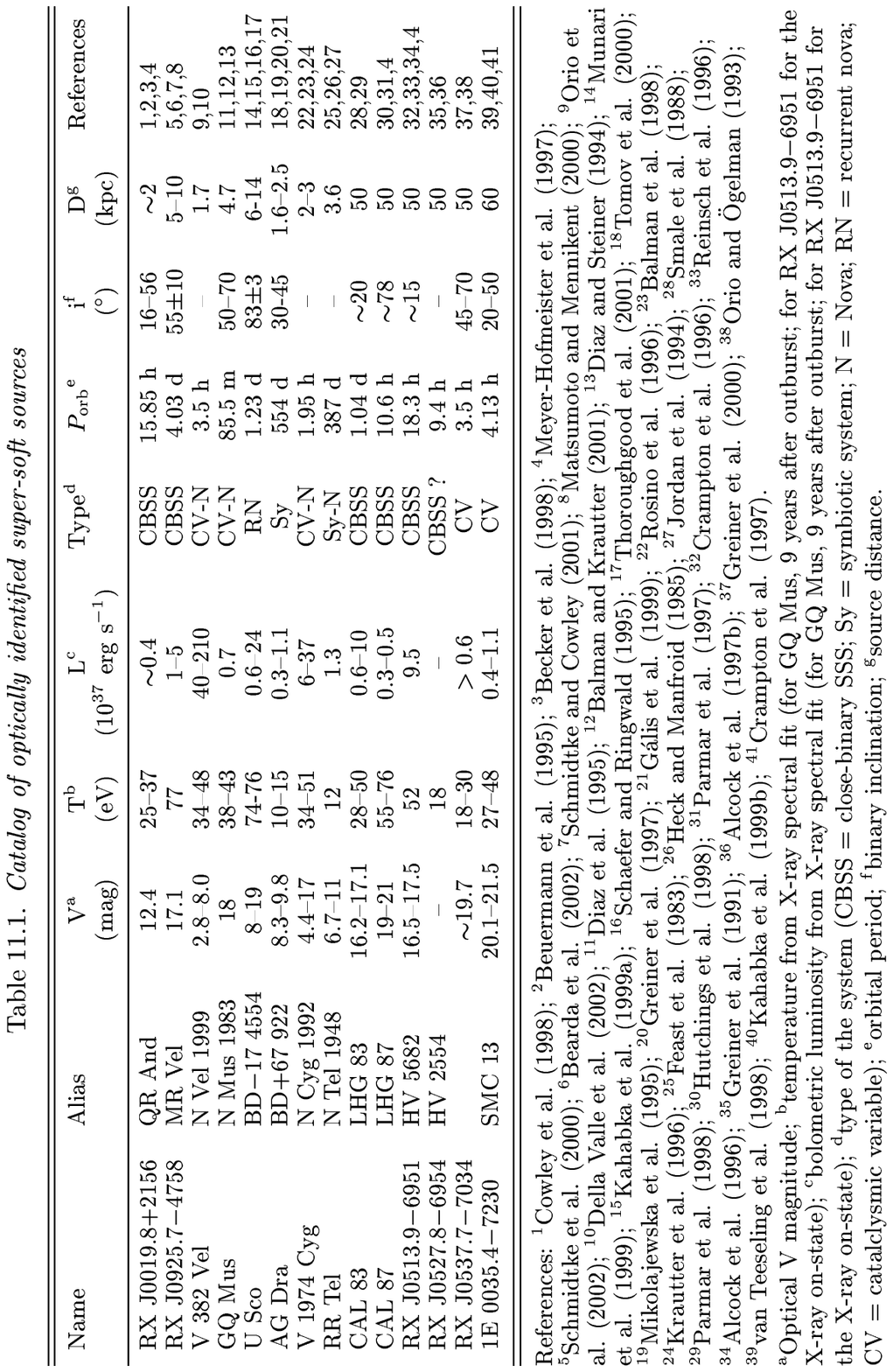,width=14.0cm,angle=-180.0,%
  bbllx=3.0cm,bblly=6.0cm,bburx=17.0cm,bbury=27.5cm,clip=}}\par
\end{figure*}

Here binary SSS are considered.
For 7 systems orbital periods of $\sim$$9^{\rm h}$ to $\sim$$4^{\rm d}$ have 
been determined, in the range predicted for steadily nuclear burning 
white dwarfs (WDs) in close-binary systems undergoing unstable mass-transfer 
on a thermal timescale (close-binary SSS, CBSS, DiStefano and Nelson 1996; 
van den Heuvel et al. 1992). From binary evolutionary calculations follows 
that such systems may have evolved companions with masses 
$\sim$$(1.3-3.0)\ M_{\odot}$ (Rappaport et al. 1994). U~Sco is a recurrent 
nova. A few systems with orbital periods of 
$1.4^{\rm h}$ to $3.5^{\rm h}$ are classical novae (CV-type SSS): V~1974 Cyg 
(Krautter et al. 1996), GQ~Mus (\"Ogelman et al. 1993), Nova LMC 1995 (Orio 
and Greiner 1999), and V~382~Vel (Orio et al. 2002). 1E~0035.4-7230 and 
RX~J0537.7$-$7034 may have a lower mass WD. The recurrent novae RS~Oph, T~CrB, 
V394~CrA, and CI Aql 2000 have a subgiant or low-mass giant donor and modeling 
of the optical outburst light curve requires a super-soft phase (Hachisu and 
Kato 2000a,b; 2001a,b). The symbiotic systems with detected super-soft X-ray 
emission RR~Tel (Jordan et al. 1994), AG~Dra (Greiner et al. 1997), SMC~3 
(Jordan et al. 1996), and Lin~358 (M\"urset et al. 1997) have orbital periods 
of at least a few hundred days (cf. Mikolajewska and Kenyon 1992). There are 
two transient SSS which have not yet been optically identified, the globular 
cluster X-ray source 1E~1339.8+2837 (Dotani et al. 1999) and RX~J0550.0$-$7151 
(Reinsch et al. 1999). For a review of SSS see Kahabka and van den Heuvel 
(1997).

\section{Nuclear burning}
Nuclear burning is ignited in an envelope of H-rich matter accreted onto a 
WD, in case a critical envelope mass $\Delta M_{\rm crit}$ has been reached 
which can sustain the high temperature ($\sim$$10^{8}\ {\rm K}$) and pressure 
($\approxgt$$(10^{18}-10^{20})\ {\rm g}\ {\rm cm^{-1}}\ {\rm s^{-1}}$)
required for nuclear burning, mainly the CNO cycle (Fujimoto 1982a,b). 
$\Delta M_{\rm crit}$ decreases with increasing WD mass $M_{\rm WD}$
and increasing accretion rate $\dot{M}_{\rm acc}$ (Prialnik and Kovetz 1995) 
and is (for a WD temperature $T_{\rm WD}$$=$$10^{7}\ {\rm K}$ and for 
$\dot{M}_{\rm acc}\ge 10^{-10}\  M_{\odot}\ {\rm yr^{-1}}$) approximated by

\begin{eqnarray*}
\log(\frac{\Delta M_{\rm crit}}{M_{\odot}}) \approx  
\end{eqnarray*}
\vskip -0.60cm
\begin{equation}
A\ + B\ \big(\frac{M_{\rm WD}}{M_\odot}\big)^{-1.436}
ln\big(1.429 - \frac{M_{\rm WD}}{M_\odot}\big) 
+ C\ \big(\log(\frac{\dot{M}_{\rm acc}}{M_{\odot}\ {\rm yr^{-1}}}) 
+ 10\big)^{1.484},
\end{equation}

\noindent
with $A = -2.862$, $B = 1.542$, and $C = -0.197$.
The accretion rate onto the WD determines the strength 
of the outburst. Higher accretion rates lead to less violent outbursts. If 
the accreted envelope remains on the WD, a steady-state can be achieved in 
case the accretion rate is similar as the nuclear burning rate. The 
steady-state accretion rate $\dot{M}_{\rm steady}$ has been given by 
Paczy\'nski and Rudak (1980) and Iben (1982). Hachisu and Kato (2001b) give 
an expression for $\dot{M}_{\rm steady}$ which is for a hydrogen content 
$X = 0.7$

\begin{equation}
\dot{M}_{\rm steady} \approx 3.7\ 10^{-7}\ 
\big(\frac{M_{\rm WD}}{M_\odot} - 0.4\big)\ M_{\odot}\ 
{\rm yr^{-1}}.
\end{equation}

\begin{figure*}[htbp]
  \vbox{\hskip -\leftskip
  \psfig{figure=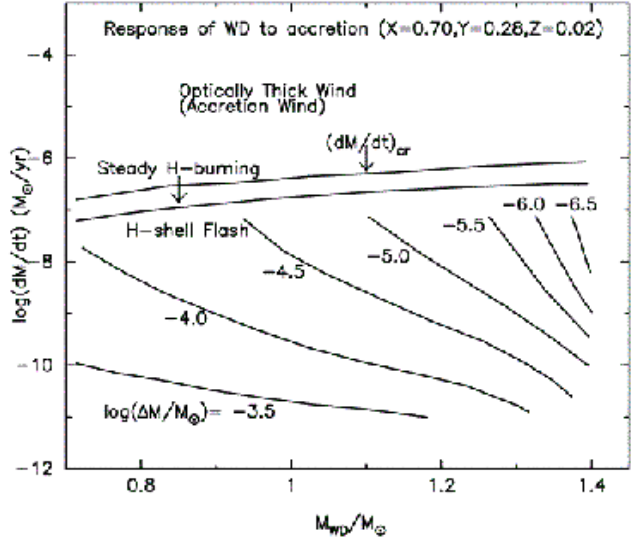,width=8.6cm,angle=0.0,%
  bbllx=4.0cm,bblly=8.1cm,bburx=17.3cm,bbury=19.6cm,clip=}}\par
  \caption{\label{fig:regimes}
  Regimes of optically thick winds, steady nuclear burning, and flashes 
  in the $M_{\rm WD}$ -- $\dot{M}_{\rm acc}$ plane (Fig.\,2 of Hachisu 
  and Kato 2001b; from Fig.\,9 of Nomoto 1982). The $\Delta M$ values 
  indicate envelope masses (for a given accretion rate) at which burning 
  is ignited.}
\end{figure*}

\noindent
$\dot{M}_{\rm steady}\ \approx 
(1 - 4)\ 10^{-7}\ {\rm M_{\odot}}\ {\rm yr^{-1}}$
for $M_{\rm WD} = (0.7 - 1.4)\ {\rm M}_{\odot}$.
For accretion rates $\dot{M}_{\rm acc} < \dot{M}_{\rm steady}$ a fraction of 
the accreted matter is ejected in a nova outburst. Below a threshold 
$\dot{M}_{\rm low} \approx 0.25\ \dot{M}_{\rm steady}$ all accreted matter 
will be ejected during a nova outburst. A red giant envelope will form for 
accretion rates above a critical rate 
$\dot{M}_{\rm crit} \approx 2\ \ \dot{M}_{\rm steady}$, 
for which part of the envelope is blown-off by a strong wind 
(cf. Fig.~{\ref{fig:regimes}}).

\section{Timescales}
In case a steady-state is sustained nuclear 
burning can last up to $\sim$$10^6$~years (the thermal timescale of the donor,
Yungelson et al. 1996). For CAL~83 an ionization nebula has been detected 
which can be explained by illumination of the local interstellar medium for 
$\sim$$10^{5}$ years by the super-soft  X-ray emission (Remillard et al. 1995).
For $\dot{M}_{\rm acc}<\dot{M}_{\rm steady}$ nuclear burning is recurrent and 
can last for $\sim$1~month to thousands of years (Prialnik and Kovetz 1995; 
Kahabka 1995; Kato 1997,1999). The timescale of expansion and contraction of 
the WD envelope due to a variable accretion rate close to 
$\dot{M}_{\rm crit}$ is given by the Kelvin-Helmholtz time
$\tau_{\rm KH} \approx 3100\ 
M_{\rm WD}\ m_{\rm env}^{-5}\ 
\big(R_{\rm WD}^{9}\ L_{\rm WD}^{37}\big)^{-1}\ {\rm days},$ 
with mass $M_{\rm WD}$ {$(M_{\odot})$}, envelope mass $m_{\rm env}^{-5}$ 
($10^{-5}$\ {$\rm M_{\odot}$), radius $R_{\rm WD}^{9}$ ($10^9\ {\rm cm}$), 
and luminosity $L_{\rm WD}^{37}$ ($10^{37}\ {\rm erg}\ {\rm s^{-1}}$) of the 
WD. There are two SSS for which alternating X-ray on and off-states have been 
discovered, RX~J0513.9$-$6951 (Pakull et al. 1993; Reinsch et al. 1996) and 
CAL~83 (Alcock et al. 1997a; Kahabka 1998; Greiner and DiStefano 2002). The 
flux pattern in both sources 
could be established in the optical due to the multi-year observational Macho 
campaign. For RX~J0513.9$-$6951 X-ray on-states of duration $\sim$$30$~days 
occur during optical dips of similar duration while X-ray off-states occur 
during optical high-states. The on/off pattern has a recurrence time of 
$\sim$$(100-200)$~days (Southwell et al. 1996). CAL~83, for which only two 
off-states have been observed, shows an irregular behavior in the optical with 
switching form high to low and intermediate states, interrupted by dipping 
with timescales of a few 10 days. A limit-cycle has been proposed for 
RX~J0513.9$-$6951 in which expansion of the WD atmosphere (due to an increase 
in the mass-accretion rate) enhances the irradiation of the accretion disk 
and the mass-flow through the disk (Reinsch et al. 2000). The involved 
timescale is the viscous time of the disk. No external mechanism is required 
to cause transitions between on and off-states. It has been proposed that the 
on/off pattern of CAL~83 is due to episodic mass-loss due to star spot 
activity of the donor (cf. Alcock et al. 1997a). An alternative explanation
is an irradiation-driven instability of the outer layer of the donor 
(cf. \v{S}imon and Mattei 1999). Other known stellar cycles are 
due to Mira type pulsations which have cycle lengths of $\sim$$(300-500)$~days.
AG~Dra has a bright giant variable with a pulsational period of 
$\sim$$355$~days (G\'alis et al. 1999), and the symbiotic nova RR~Tel has an 
AGB Mira variable with a period of $\sim$$(350-410)$~days (Heck and Manfroid 
1985). The similarity in the optical variability of several unusual CVs, 
V~Sagittae, T~Pyxidis, WX~Centauri, and V~751~Cygni, with RX~J0513.9$-$6951, 
let to propose that these systems belong to the class of SSS (Patterson et al. 
1998; Greiner 2000b).

\section{Super-soft X-ray spectra} 
The maximum atmospheric temperature $T_{\rm max}$ of steadily nuclear burning 
cold ($<$$10^{7}\ {\rm K}$) WDs which have cooling ages 
$\approxgt (1-3)\ 10^{8}\ {\rm yr}$ (Yungelson et al. 1996) is for
$M_{\rm WD} \sim$$(0.7 - 1.3)\ M_{\odot}$ (from Iben 1982)

\begin{equation}
T_{\rm max} \approx
1.4\ 10^{6}\ \big(0.107 
+ \big(\frac{M_{\rm WD}}{M_{\odot}} - 0.6\big)^{1.7} \big) \ {\rm K}.
\end{equation}

\noindent
Sion and Starrfield (1994) have calculated evolutionary models for low-mass 
($M_{\rm WD}<0.7\ M_{\odot}$) hot WDs accreting at rates 
$\sim$$10^{-8}\ M_{\odot}\ {\rm yr^{-1}}$ over thousands of years. The WD is 
heated up during the evolution due to steady nuclear burning and reaches a 
surface temperature $T_{\rm eff}\sim$$3.25\ 10^{5}\ {\rm K}$. MacDonald 
(1996) gives the maximum temperature for a WD after a nova outburst and before 
turn-off. The WD mass can also be estimated from the bolometric luminosity 
during the plateau phase of the Hertzsprung-Russel diagram (Iben 1982).
The temperature is a better indicator of the WD mass as the luminosity can be 
reduced due to absorption and scattering. Effective temperatures in the range 
$\sim$20 to $\sim$80~eV and bolometric luminosities 
$\sim$$3\ 10^{36}$ to $\sim$$3\ 10^{38}\ {\rm erg}\ {\rm s^{-1}}$ 
(cf. Tab.~11.1), have been derived by fitting blackbody, LTE 
and non-LTE WD model atmospheres (e.g. Hartmann and Heise 1997) to the X-ray 
spectra of SSS measured with {\sl ROSAT} (van Teeseling et al. 1994; Balman 
et al. 1998), {\sl ASCA} (Asai et al. 1998; 
Ebisawa et al. 2001) and {\sl BeppoSAX} (Parmar et al. 1997,1998; Hartmann et 
al. 1999; Kahabka et al. 1999a,b; Orio et al. 2002). The highly resolved 
spectrum of CAL~83 measured with {\sl XMM-Newton} (Paerels et al. 2001)
is dominated by numerous absorption or emission features. The {\sl Chandra} 
(Bearda et al. 2002) and {\sl XMM-Newton} (Motch et al. 2002) spectra of 
RX~J0925.7$-$4758 show a wealth of spectral features with emission 
lines of highly ionized metals (e.g. \ion{O}{viii} and \ion{Fe}{xvii}) which 
show P~Cygni wind profiles. For a few SSS the neutral hydrogen absorbing column
which is required for the spectral modeling is available from UV measurements 
(G\"ansicke et al. 1998). 

\section{The accretion disk}
Super-soft X-ray emission is reprocessed by the accretion disk into UV and 
optical light (Fukue and Matsumoto 2001; Suleimanov et al. 1999). Modeling of 
the optical orbital light curve of CAL~87, RX~J0019.8+2156, 1E~0035.4$-$7230, 
and CAL~83 has been performed by Schandl et al. (1997); Meyer-Hofmeister et 
al. (1997); Kitabatake and Fukue (2002). A variation of the disk rim height 
with orbital phase is explained by the intersection of the gas stream with the 
accretion disk which causes a bulge (spray) on the disk rim. For the 
\ion{He}{ii}$~\lambda4686$, $\rm H\alpha$, and $\rm H\beta$ emission lines 
blue- and redshifted satellites (bipolar jets) have been found in the optical 
spectra of RX~J0019.8+2156, RX~J0513.9$-$6951, and RX~J0925.7$-$4758, which 
correspond to Doppler velocities $v_{\rm bipolar}$ $\sim$$885$, $\sim$$3800$, 
and $\sim$$5200$ ${\rm km}\ {\rm s}^{-1}$ (Tomov et al. 1998; Becker et al. 
1998; Southwell et al. 1996; Motch et al. 1998). These lines are transient 
on timescales of months. One can compare these velocities with the terminal 
velocity of a line driven wind, $v_{\infty}^{ld} = 3\ v_{\rm esc}$, with the 
escape velocity
 $v_{\rm esc} \approx 5160\ \sqrt {M_{\rm WD}(M_{\odot}) /  R_{\rm d}(10^9\ 
{\rm cm})} \ {\rm km}\ {\rm s^{-1}}$ (Cassinelli 1979) 
and of a radiation driven wind from an accretion disk surrounding a WD, 
$v_{\infty}^{rd} = 0.42\  v_{\rm esc}$ (Hachiya et al. 1998; Fukue 
and Hachiya 1999). In luminous SSS the gaseous envelope is highly ionized and 
radiation driven winds are dominating. For $M_{\rm WD}\approx 0.6$ to 
$1.4 M_{\odot}$ one derives in case of radiation driven winds (from the inner 
disk with radius $R_{\rm d}$) terminal velocities of 1700 to 5700 
${\rm km}\ {\rm s}^{-1}$, consistent with the observed velocities of bipolar 
outflows. Bearda et al. (2002) found in the complex {\sl Chandra} {\sl HETGS} 
spectrum of RX~J0925.7$-$4758 P~Cygni structure in the line profiles of 
\ion{O}{viii} (Ly~$\beta$) and \ion{Fe}{xvii}. The highest absorption 
velocities of $\sim$$1500\ {\rm km}\ {\rm s^{-1}}$ are consistent with 
radiation driven winds from the inner disk of a massive WD.

\section{The donor star}
In CBSS the donor is assumed to be slightly evolved, between the zero-age 
main sequence and the terminal-age main sequence (TAMS) or even beyond
(subgiant), and to nearly fill its Roche lobe. For stars at the TAMS with 
$M_{\rm don}\approxgt1.2\ M_{\odot}$, $<$$\rho$$>$$\approx20.9\ 
(10^{M_{\rm don}})^{-1.54}+0.0457\ (10^{M_{\rm don}})^{-0.045}\ 
{\rm g}\ {\rm cm^{-3}}$, assuming a mass-radius relation for solar metallicity 
stars (Schaller et al. 1992). For CAL~83, assuming a Roche lobe filling donor 
and $M_{\rm WD}\approx1.0\ M_{\odot}$, the mean density is
$<$$\rho$$>$$\approx0.15\ {\rm g}\ {\rm cm^{-3}}$, consistent with 
$M_{\rm don}\approxgt1.5\ M_{\odot}$ below the TAMS (cf. van den Heuvel et al. 
1992). In case 
a significant amount of matter has been transferred during the evolution of 
the system this mass estimate may not be valid. A commonly used method to 
constrain $M_{\rm don}$  is by radial velocity emission line diagnostics of 
the \ion{He}{ii}$\lambda$4686 line, assuming the line originates predominantly 
in the accretion disk (Thoroughgood et al. 2001; Matsumoto and Mennickent 2000;
Becker et al. 1998). In case the \ion{He}{ii}$\lambda$4686 line emission is 
strongly dominated by the mass-flow from the donor (cf. Cowley et al. 2002), 
then the inferred donor mass can be underestimated. The location of the 
\ion{He}{ii}$\lambda$4686 line in the binary system can be inferred from 
Doppler maps (cf. Deufel et al. 1999). The optical spectra of CBSS are 
dominated by the emission of the bright accretion disk. The donor has a minor 
contribution to the optical spectrum ($\sim$10\%), the spectral type can in 
most cases not be determined. An exception are SSS with optical faint states 
like recurrent novae: The accretion disk has a minor contribution to the 
overall spectrum, irradiation is strongly reduced, and the donor dominates the 
optical spectrum. Pritchet and van den Bergh (1977) give for spectral features 
\ion{Mg}{i}+Mg~H, \ion{Fe}{i}+\ion{Ca}{i}, \ion{Na}{i}+TiO the line equivalent 
width as a function of spectral type (for giant and main-sequence stars). 
Anupama and Dewangan (2000) have measured the 
\ion{Mg}{i}b and \ion{Fe}{i}+\ion{Ca}{i} absorption features in the spectrum 
of the 1999 outburst of U~Sco and have determined the spectral type as K2~V. 
Matsumoto and Mennickent (2000) constrained from the non-detection of 
\ion{Si}{iii} $\lambda$4525 and \ion{Mg}{ii} $\lambda$4481 in the optical 
spectrum of RX~J0925.7$-$4758 the spectral type of the donor. For an 
inclination $i\sim$50\D, $M_{\rm don}\sim$$3.5\ M_{\odot}$, 
lower than for a nearly Roche-lobe filling solar metallicity star at the TAMS, 
but consistent with a subgiant.

\section{Evolution}
The evolution of CBSS undergoing unstable mass-transfer 
($M_{\rm don}$$>$$M_{\rm WD}$) has been considered by Rappaport et al. (1994); 
Yungelson et al. (1996), and Langer et al. (2000). An estimate of 
the mean mass-transfer rate in CBSS is 
$\dot{M} \approx M_{\rm don} / t_{\rm th}$, with 
$t_{\rm th} = 3\ 10^{7}\ (M_{\rm don}/M_{\odot})^{-2}\ {\rm yr}$,
the thermal timescale of the donor (van den Heuvel 1994). For initial donor 
masses $M_{\rm don}\sim$$(1.5-2.5)\ M_{\odot}$ mass-transfer rates 
$\dot{M} \approx(1-5)\ 10^{-7}\ M_{\odot}\ {\rm yr}^{-1}$ 
are derived. There have been proposed alternative evolutionary scenarii for 
CBSS, i.e. irradiation-driven mass-transfer (van Teeseling and 
King 1998) which could account for systems in the short period 
($\approxlt$$4^{\rm h}$) regime (see also Ritter 2000). 
CO WDs which have masses $<$$1.2\ M_{\odot}$ and which grow in mass 
by accretion towards the Chandrasekhar mass ($\sim$$1.38\ M_{\odot}$) have 
been proposed as candidate progenitors for Type Ia supernovae. The large 
accretion rates required for steady nuclear burning can be supplied either 
by slightly evolved main sequence stars or by low-mass red giants. The 
realization frequency can be as large as $(1-3)\ 10^{-3}\ {\rm yr^{-1}}$ and 
close to the galactic supernova Ia rate of $\sim$$3\ 10^{-3}\ {\rm yr^{-1}}$ 
(Nomoto et al. 2000; Hachisu et al. 1996; Li and van den Heuvel 1997; 
Yungelson and Livio 1998).

\section{SSS in nearby galaxies}
During the {\sl ROSAT} survey of M31 15 SSS and the 
recurrent super-soft transient RX~J0045.4+4154 have been detected (Supper et 
al. 1997; White et al. 1995). One transient SSS (RX~J0044.0+4118) 
has been optically identified with a classical nova in M31 (Nedialkov et al. 
2002). The detected sources have luminosities 
$\approxgt 2\ 10^{37}\ {\rm erg}\ {\rm s^{-1}}$ and may be connected to 
massive ($\approxgt 0.9\ M_{\odot}$) steadily nuclear burning WDs 
(Kahabka 1999). During recent surveys performed with {\sl Chandra} and 
{\sl XMM-Newton} SSS were detected to considerably lower 
luminosities of $\sim$$(2-8)\ 10^{36}\ {\rm erg}\ {\rm s^{-1}}$. Two sources, 
XMM~J004308.5+411820 (Shirey 2001) and XMM~J004414.0+412204 (Trudolyubov et al.
2002), are transients. In addition the first super-soft pulsator 
(865~s) has been discovered with {\sl XMM-Newton} (XMM~J004319.4+411759), which
is transient and more luminous (Osborne et al. 2001; Trudolyubov et 
al. 2001). This period is below the orbital period of accreting 
main-sequence star binaries. An explanation as a double-degenerate system is 
unlikely, but a super-soft intermediate polar (M31PSS) appears to be 
possible (King et al. 2002). 3 of the 5 SSS observed after {\sl ROSAT} are 
transients. Such a high fraction of transient SSS is consistent with the 
numbers derived from population synthesis calculations (Yungelson et al. 1996).
In M33 7 SSS were detected 
(Haberl and Pietsch 2001). 8 SSS have been discovered with 
{\sl Chandra} in the Sb spiral galaxy M~81, distance 3.6~Mpc (Swartz et al. 
2002). The sources have bolometric luminosities of $\sim$$2\ 10^{36}$ to 
$\approxgt$$3\ 10^{38}\ {\rm erg}\ {\rm s}^{-1}$. 4 of the sources are 
located in the bulge and 3 sources coincide with spiral arms. The brightest 
source (N1) is a bulge source with a luminosity 
$\approxgt$$3\ 10^{38}\ {\rm erg}\ {\rm s}^{-1}$ and effective temperature 
$\sim$$(70-90)$~eV. The large luminosity would require a massive WD, in a 
state of luminous outflow. From the age of the bulge population ($>$$8$~Gyr), 
a mass $<$$1\ M_{\odot}$ is inferred for the stellar companion. The system 
could be a helium Algol (Iben 
and Tutukov 1994). Two further bright sources N2 and N3 with effective 
temperatures $(7.9 - 8.5)\ 10^{5}\ {\rm K}$ and $(5.4 - 7.5)\ 10^{5}\ {\rm K}$ 
are consistent with massive WDs. 10 SSS have 
been discovered with {\sl Chandra} in the Scd spiral galaxy M101, distance 
7.2~Mpc (Pence et al. 2001). The 3 brighter sources have luminosities 
$\sim$$10^{38}\ {\rm erg}\ {\rm s}^{-1}$ and effective temperatures 
$\sim$$(7 - 11)\ 10^{5}$~K. 7 fainter sources have effective temperatures
$\sim$$(5.5\pm0.2)\ 10^{5}$~K. Most SSS correlate with 
spiral arms, one source is located in an interarm region, and another source 
correlates within 2\arcsec\ with a globular cluster. In NGC~55, NGC~300, and 
NGC~1291 one SSS has been detected (Schlegel et al. 1997; Read and Pietsch 
2001; Irwin et al. 2002). Sarazin et al. (2001) discovered with {\sl Chandra} 
the first three SSS in an elliptical galaxy, NGC~4697. At a distance of 
$\sim$$16$~Mpc these sources must be luminous 
($\approxgt5\ 10^{37}\ {\rm erg}\ {\rm s^{-1}}$) to be detectable. To evolve 
towards the TAMS within the age of the elliptical a stellar 
mass $\sim$$1.0\ M_{\odot}$ is required, below the lower limit mass for CBSS. 
These sources may be CV-type SSS.

\end{document}